\documentclass[iop]{emulateapj}
\pdfoutput=1
\usepackage{graphicx, subfigure}
\usepackage{morefloats}
\usepackage{amsmath}
\usepackage{amssymb}
\usepackage{mathrsfs,amssymb}
\usepackage{url}
\usepackage{color}

\usepackage{mwe,tikz}
\usepackage[percent]{overpic}

\usepackage{rotating}

\usepackage{floatrow}

\usepackage{natbib}

\usepackage{letltxmacro}

\LetLtxMacro{\ORIcitep}{\citep}
\DeclareRobustCommand{\citep}{\leavevmode\unskip~\ORIcitep}

\begin{document}

\title{{Synchrotron Origin of the Typical GRB Band Function \\ --  A Case Study of GRB 130606B}}

 
\author{Bin-Bin Zhang\altaffilmark{1,*}, Z. Lucas Uhm\altaffilmark{2}, Valerie Connaughton\altaffilmark{3},Michael S. Briggs\altaffilmark{1}, Bing Zhang\altaffilmark{2}}

\altaffiltext{*}{Contact Email: binbin.zhang@uah.edu}

\affil{$^{1}$Center for Space Plasma and Aeronomic Research (CSPAR), University
of Alabama in Huntsville, Huntsville, AL 35899, USA} 
\affil{$^{2}$Department of Physics and Astronomy, University of Nevada, Las Vegas, NV 89154, USA}
\affil{$^{3}$Science and Technology Institute, Universities Space Research Association, Huntsville, AL 35805, USA}

\begin{abstract}

We perform a time-resolved spectral analysis of GRB 130606B within the framework of a fast-cooling synchrotron radiation model with magnetic field strength in the emission region decaying with time, as proposed by Uhm \& Zhang. The data from all time intervals can be successfully fit by the model. The same data can be equally well fit by the empirical Band function with typical parameter values. Our results, which involve only minimal physical assumptions, offer one natural solution to the origin of the observed GRB spectra and imply that, {at least some, if not all, Band-like GRB spectra with typical Band parameter values can indeed be explained by synchrotron radiation.}
 \end{abstract}

\section{Introduction}

Despite the multifarious observations over 40 years since {their discovery, the prompt emission phase of Gamma-Ray Bursts (GRBs) is still not fully understood} in several fundamental aspects (e.g., \citealt{zhangb11_review_zbb} for a review): (1) What is the composition of the ejecta (baryonic fireball or Poynting flux dominated jet)? (2) What are the energy dissipation and particle acceleration mechanisms (from shocks or via magnetic reconnections)? (3) What is the radiation mechanism (synchrotron, synchrotron self-Compton, or Comptonization of quasi-thermal emission from the photosphere)? {Interpretation of the observed GRB energy spectra plays a key role to answer these questions.} A direct identification of the GRB prompt emission mechanism would lend further constraints on the energy dissipation mechanism, and eventually on the composition of the GRB jets.

GRB spectra have traditionally been described by a joint smoothly broken power law function - the ``Band" function\citep{band93_zbb}. This function successfully captures the main features of the GRB spectra\citep[e.g.,][]{briggs99_zbb,abdo09_zbb_080916c,zhangbb11_zbb} but it has been difficult to associate the parameters with the physical processes responsible for the GRB emission. 
Two leading candidate radiation mechanisms for GRB prompt emission are
synchrotron radiation\citep{meszaros94_zbb,Daigne98_Sync_zbb,lloyd00_zbb,zhangyan11_zbb,Daigne11_sync_zbb} and Comptonization of quasi-thermal photons from the photosphere of the fireball\citep{thompson94_zbb,rees05_zbb,beloborodov10_zbb,lazzati10_thermal_zbb}, which both {successfully} reproduce the observed values of $E_p$ (peak energy in the $\nu F_\nu$ spectrum) and $\beta$ (the high-energy power-law spectral index). The straightforward predictions of the low-energy power-law index $\alpha$ of both models, however, deviate from the typical value $\alpha \sim -1$ revealed by the observations\citep[e.g.,][]{Preece00_batsecatlog_zbb,zhangbb11_zbb,Nava11_MN_zbb}. The traditional fast-cooling synchrotron model predicts an $\alpha$ value of $-1.5$\citep{sari98_zbb}. This has been regarded as a strong drawback of the traditional fast-cooling synchrotron model to interpret GRB prompt emission\citep{Ghisellini_00_constrains_zbb}. The photosphere models, on the other hand, predict too hard {on} spectrum. A pure blackbody emission in Rayleigh-Jeans regime predicts $\alpha$ $\sim$ +1. Relativistic effects (equal-arrival-time surface effect) would soften it to $\alpha$ $\sim +0.4$\citep[e.g.,][]{beloborodov10_zbb,deng14_zbb_ep}, which is still too hard compared to the data. Effects that modify the low-energy photon index closer to -1 include adding synchrotron self-absorption or a smooth cutoff in the low-energy electron distribution\citep{lloyd00_zbb}, adding a photosphere component in the synchrotron model\citep{meszaros10_zbb_steepslope}, adding a synchrotron component in the photosphere model\citep{toma11_zbb,vurm11_zbb}, or introducing a certain type of structured jet\citep{lundman13_zbb}. However, some of these modifications either invoke {simplified} assumptions or introduce spectral features that are inconsistent with the observations.

\begin{sidewaystable}
\setlength{\tabcolsep}{2pt}

\begin{tiny}
\begin{center}

\begin{tabular}{lllllllllllllll}

\hline
Interval & t1 $\sim$ t2 (s) & $\alpha$ & $\beta$ & $E_p$ (keV) & PGSTAT/dof & log $\Gamma$ & $p$ & log $\gamma_{inj}$ & log R$_{inj}^0$ (s$^{-1}$) & q & log B$_0$ (G) & b & $\hat{t}$ (s) & PGSTAT/dof \\ 
\hline
\hline
0&0.00 $\sim$ 2.94 &$-0.52_{-0.24}^{+0.08}$& $-1.89_{-0.45}^{+0.04}$&$428.89_{-40.40}^{+306.11}$& 325.71/360&$2.82_{-0.40}^{+0.03}$&$3.30_{-0.15}^{+1.00}$&$5.52_{-0.17}^{+0.17}$&$45.73_{-0.50}^{+0.50}$&$3.32_{-1.67}^{+0.66}$&$1.02_{-0.05}^{+1.05}$&$0.98_{-0.08}^{+0.31}$&$1.51_{-0.96}^{+1.77}$&326.44/356\\
1&2.94 $\sim$ 5.52 &$-0.84_{-0.07}^{+0.02}$& $-2.59_{-6.28}^{+0.53}$&$611.01_{-5.66}^{+127.25}$& 319.18/360&$2.49_{-0.21}^{+0.29}$&$4.58_{-0.44}^{+1.02}$&$5.52_{-0.13}^{+0.13}$&$44.87_{-0.60}^{+0.24}$&$3.41_{-1.02}^{+1.02}$&$1.12_{-0.14}^{+0.91}$&$0.98_{-0.08}^{+0.31}$&$3.12_{-2.18}^{+2.26}$&314.84/356\\
2&5.52 $\sim$ 8.10 &$-0.65_{-0.09}^{+0.05}$& $-2.15_{-0.22}^{+0.05}$&$309.92_{-16.91}^{+46.78}$& 305.41/360&$2.63_{-0.22}^{+0.24}$&$3.91_{-0.12}^{+0.93}$&$5.92_{-0.16}^{+0.08}$&$45.55_{-0.54}^{+0.54}$&$4.81_{-1.04}^{+0.19}$&$1.03_{-0.15}^{+0.81}$&$1.71_{-0.56}^{+0.05}$&$3.52_{-0.42}^{+3.59}$&308.83/356\\
3&8.10 $\sim$ 9.55 &$-0.88_{-0.03}^{+0.02}$& $-1.97_{-0.05}^{+0.04}$&$683.67_{-46.04}^{+67.58}$& 398.80/360&$2.17_{-0.16}^{+0.26}$&$2.58_{-0.14}^{+0.14}$&$5.62_{-0.10}^{+0.16}$&$45.38_{-0.99}^{+0.16}$&$2.68_{-1.56}^{+0.62}$&$0.82_{-0.12}^{+0.58}$&$1.19_{-0.05}^{+0.59}$&$5.13_{-0.33}^{+3.54}$&385.33/356\\
4&9.55 $\sim$ 10.79 &$-0.84_{-0.02}^{+0.02}$& $-2.11_{-0.06}^{+0.05}$&$638.50_{-38.64}^{+46.45}$& 391.76/360&$2.14_{-0.14}^{+0.26}$&$3.04_{-0.01}^{+0.29}$&$5.64_{-0.06}^{+0.17}$&$44.78_{-0.45}^{+0.73}$&$4.10_{-0.99}^{+0.90}$&$0.84_{-0.14}^{+0.60}$&$1.15_{-0.03}^{+0.56}$&$5.84_{-0.88}^{+2.93}$&375.45/356\\
5&10.79 $\sim$ 12.42 &$-0.84_{-0.03}^{+0.03}$& $-2.30_{-0.15}^{+0.06}$&$473.64_{-27.74}^{+38.34}$& 390.44/360&$2.15_{-0.15}^{+0.28}$&$3.53_{-0.28}^{+0.28}$&$5.62_{-0.06}^{+0.22}$&$44.09_{-0.09}^{+0.56}$&$4.06_{-1.02}^{+0.91}$&$0.86_{-0.16}^{+0.53}$&$1.11_{-0.03}^{+0.59}$&$7.78_{-3.00}^{+1.04}$&373.70/356\\
6&12.42 $\sim$ 13.67 &$-0.92_{-0.03}^{+0.02}$& $-1.97_{-0.05}^{+0.03}$&$844.39_{-51.17}^{+76.54}$& 411.70/360&$2.54_{-0.30}^{+0.19}$&$2.81_{-0.26}^{+0.01}$&$5.84_{-0.25}^{+0.00}$&$46.05_{-0.61}^{+0.45}$&$0.26_{-0.23}^{+0.87}$&$1.67_{-0.59}^{+0.67}$&$1.75_{-0.54}^{+0.06}$&$4.85_{-0.53}^{+3.94}$&413.08/356\\
7&13.67 $\sim$ 14.89 &$-1.09_{-0.02}^{+0.02}$& $-1.88_{-0.05}^{+0.03}$&$1037.46_{-82.16}^{+127.90}$& 361.48/360&$2.37_{-0.24}^{+0.27}$&$2.20_{-0.11}^{+0.09}$&$5.63_{-0.13}^{+0.10}$&$45.87_{-0.93}^{+0.18}$&$0.09_{-0.08}^{+0.61}$&$2.07_{-0.86}^{+0.48}$&$1.88_{-0.33}^{+0.12}$&$5.40_{-1.53}^{+3.41}$&367.51/356\\
8&14.89 $\sim$ 16.62 &$-1.31_{-0.02}^{+0.01}$& $-2.44_{-6.07}^{+0.40}$&$1343.87_{-28.36}^{+338.42}$& 391.27/360&$2.18_{-0.08}^{+0.32}$&$5.10_{-1.26}^{+0.46}$&$5.71_{-0.13}^{+0.05}$&$45.88_{-0.43}^{+0.43}$&$0.07_{-0.06}^{+0.44}$&$2.02_{-0.40}^{+0.64}$&$1.95_{-0.31}^{+0.05}$&$5.60_{-1.33}^{+3.32}$&408.31/356\\
9&16.62 $\sim$ 18.93 &$-1.42_{-0.02}^{+0.02}$& $-13.85_{-3.46}^{+8.02}$&$592.61_{-54.26}^{+91.16}$& 331.80/360&$2.14_{-0.06}^{+0.36}$&$5.88_{-0.65}^{+0.11}$&$5.44_{-0.10}^{+0.10}$&$45.67_{-0.55}^{+0.55}$&$0.14_{-0.13}^{+1.11}$&$1.79_{-0.40}^{+0.42}$&$1.07_{-0.06}^{+0.63}$&$8.13_{-3.62}^{+0.57}$&336.67/356\\
10&18.93 $\sim$ 21.78 &$-1.63_{-0.05}^{+0.04}$& $-17.22_{-0.12}^{+11.71}$&$148.23_{-15.46}^{+27.29}$& 278.08/360&$2.07_{-0.02}^{+0.22}$&$3.37_{-0.33}^{+1.62}$&$4.56_{-0.17}^{+0.18}$&$46.05_{-0.62}^{+0.62}$&$0.53_{-0.52}^{+1.12}$&$1.98_{-0.32}^{+0.32}$&$1.11_{-0.11}^{+0.45}$&$2.31_{-1.46}^{+2.28}$&279.55/356\\
11&21.78 $\sim$ 25.00 &$-1.29_{-0.13}^{+0.13}$& $-2.26_{-7.03}^{+7.03}$&$39.84_{-11.06}^{+11.06}$& 306.47/360&$2.22_{-0.15}^{+0.31}$&$2.79_{-0.03}^{+0.84}$&$4.64_{-0.36}^{+0.12}$&$44.46_{-0.45}^{+0.58}$&$1.60_{-1.03}^{+1.54}$&$2.01_{-0.40}^{+0.69}$&$1.22_{-0.16}^{+0.48}$&$8.64_{-2.15}^{+1.36}$&306.52/356\\
12&25.00 $\sim$ 28.19 &$-1.53_{-0.17}^{+0.08}$& $-2.80_{-6.46}^{+6.46}$&$54.49_{-3.42}^{+13.09}$& 285.05/360&$2.47_{-0.33}^{+0.20}$&$3.42_{-0.18}^{+1.11}$&$4.81_{-0.38}^{+0.33}$&$44.61_{-0.18}^{+0.79}$&$1.38_{-0.79}^{+1.12}$&$2.21_{-0.92}^{+0.44}$&$1.60_{-0.54}^{+0.11}$&$6.43_{-3.10}^{+1.84}$&284.06/356\\
13&28.19 $\sim$ 31.16 &$-1.49_{-0.07}^{+0.04}$& $-15.29_{-2.03}^{+9.82}$&$294.85_{-35.03}^{+131.17}$& 313.87/360&$2.20_{-0.15}^{+0.14}$&$5.71_{-0.96}^{+0.29}$&$4.86_{-0.17}^{+0.17}$&$45.22_{-0.83}^{+0.27}$&$1.76_{-1.24}^{+1.05}$&$2.28_{-0.07}^{+0.58}$&$1.25_{-0.25}^{+0.38}$&$3.02_{-0.16}^{+5.28}$&314.13/356\\
14&31.16 $\sim$ 34.18 &$-1.43_{-0.10}^{+0.07}$& $-16.97_{-0.28}^{+11.73}$&$130.97_{-13.65}^{+31.40}$& 337.20/360&$2.87_{-0.20}^{+0.13}$&$5.32_{-2.02}^{+0.12}$&$4.71_{-0.09}^{+0.27}$&$44.15_{-0.15}^{+0.49}$&$0.16_{-0.15}^{+1.00}$&$2.64_{-0.83}^{+0.11}$&$0.96_{-0.06}^{+0.33}$&$3.29_{-0.45}^{+4.70}$&337.13/356\\
15&34.18 $\sim$ 37.18 &$-1.32_{-0.07}^{+0.04}$& $-2.90_{-6.16}^{+0.95}$&$298.60_{-20.85}^{+88.30}$& 306.05/360&$2.42_{-0.32}^{+0.10}$&$5.31_{-1.86}^{+0.19}$&$5.32_{-0.12}^{+0.12}$&$44.74_{-0.46}^{+0.41}$&$0.12_{-0.11}^{+0.73}$&$2.46_{-0.83}^{+0.13}$&$1.41_{-0.33}^{+0.31}$&$9.66_{-2.49}^{+0.34}$&308.32/356\\
16&37.18 $\sim$ 39.45 &$-0.96_{-0.03}^{+0.03}$& $-18.84_{-1.16}^{+5.68}$&$433.62_{-22.96}^{+27.09}$& 394.86/360&$2.72_{-0.25}^{+0.25}$&$5.91_{-0.46}^{+0.09}$&$5.74_{-0.21}^{+0.02}$&$45.06_{-0.54}^{+0.52}$&$0.37_{-0.36}^{+0.80}$&$2.68_{-0.67}^{+0.31}$&$1.76_{-0.57}^{+0.04}$&$8.71_{-4.16}^{+0.00}$&399.39/356\\
17&39.45 $\sim$ 41.55 &$-0.87_{-0.05}^{+0.04}$& $-2.31_{-0.27}^{+0.08}$&$282.59_{-15.57}^{+34.85}$& 332.79/360&$2.36_{-0.03}^{+0.48}$&$3.68_{-0.00}^{+0.77}$&$5.53_{-0.08}^{+0.29}$&$44.47_{-0.00}^{+1.16}$&$2.72_{-0.84}^{+0.84}$&$1.09_{-0.06}^{+1.21}$&$1.17_{-0.03}^{+0.64}$&$7.88_{-3.17}^{+0.89}$&326.97/356\\
18&41.55 $\sim$ 44.14 &$-0.95_{-0.07}^{+0.01}$& $-2.81_{-6.60}^{+0.61}$&$192.70_{-11.76}^{+11.76}$& 260.84/360&$2.13_{-0.13}^{+0.25}$&$5.05_{-0.65}^{+0.54}$&$5.36_{-0.07}^{+0.29}$&$44.85_{-0.44}^{+0.68}$&$2.81_{-2.09}^{+0.19}$&$0.86_{-0.15}^{+0.56}$&$1.01_{-0.10}^{+0.29}$&$5.45_{-1.52}^{+2.93}$&258.88/356\\
19&44.14 $\sim$ 46.84 &$-0.95_{-0.05}^{+0.05}$& $-13.78_{-3.61}^{+7.42}$&$166.12_{-7.49}^{+9.45}$& 311.29/360&$2.21_{-0.21}^{+0.25}$&$5.85_{-0.66}^{+0.02}$&$5.39_{-0.07}^{+0.29}$&$44.24_{-0.23}^{+0.56}$&$3.21_{-0.87}^{+0.87}$&$1.54_{-0.58}^{+0.56}$&$1.84_{-0.29}^{+0.16}$&$7.24_{-2.73}^{+1.42}$&315.96/356\\
20&46.84 $\sim$ 49.72 &$-0.71_{-0.07}^{+0.07}$& $-2.55_{-6.60}^{+0.44}$&$126.27_{-10.47}^{+10.47}$& 331.01/360&$2.89_{-0.23}^{+0.11}$&$5.59_{-0.73}^{+0.17}$&$5.71_{-0.33}^{+0.07}$&$44.05_{-0.05}^{+0.49}$&$4.91_{-0.95}^{+0.09}$&$1.69_{-0.76}^{+0.23}$&$1.51_{-0.31}^{+0.28}$&$5.80_{-1.51}^{+1.95}$&329.55/356\\
21&49.72 $\sim$ 51.63 &$-0.75_{-0.07}^{+0.02}$& $-2.51_{-1.43}^{+0.06}$&$284.99_{-8.27}^{+40.89}$& 365.09/360&$2.78_{-0.43}^{+0.03}$&$4.97_{-0.16}^{+0.68}$&$5.93_{-0.14}^{+0.07}$&$44.57_{-0.30}^{+0.61}$&$3.59_{-1.13}^{+0.37}$&$1.70_{-0.76}^{+0.46}$&$1.60_{-0.36}^{+0.26}$&$7.40_{-1.93}^{+1.31}$&357.80/356\\
22&51.63 $\sim$ 53.63 &$-0.71_{-0.07}^{+0.01}$& $-2.90_{-6.44}^{+0.63}$&$241.52_{-0.22}^{+25.16}$& 350.36/360&$2.87_{-0.39}^{+0.00}$&$5.99_{-0.26}^{+0.01}$&$5.91_{-0.25}^{+0.01}$&$44.25_{-0.04}^{+0.74}$&$4.76_{-0.68}^{+0.24}$&$2.28_{-0.53}^{+0.53}$&$1.84_{-0.29}^{+0.15}$&$6.33_{-1.16}^{+1.24}$&348.97/356\\
23&53.63 $\sim$ 56.16 &$-0.93_{-0.04}^{+0.04}$& $-3.83_{-5.93}^{+1.49}$&$217.80_{-9.21}^{+13.21}$& 340.78/360&$2.00_{-0.00}^{+0.25}$&$5.90_{-0.48}^{+0.10}$&$5.58_{-0.19}^{+0.17}$&$44.46_{-0.06}^{+0.98}$&$2.70_{-1.70}^{+0.14}$&$0.87_{-0.17}^{+0.51}$&$1.58_{-0.42}^{+0.21}$&$9.80_{-1.92}^{+0.19}$&338.91/356\\
24&56.16 $\sim$ 58.97 &$-0.88_{-0.06}^{+0.07}$& $-3.76_{-6.01}^{+1.49}$&$133.90_{-6.86}^{+8.80}$& 273.32/360&$2.83_{-0.47}^{+0.01}$&$6.00_{-0.40}^{+0.00}$&$5.61_{-0.27}^{+0.21}$&$46.11_{-0.54}^{+0.44}$&$2.69_{-1.07}^{+0.99}$&$1.06_{-0.12}^{+0.99}$&$1.42_{-0.08}^{+0.42}$&$2.37_{-1.52}^{+2.15}$&271.40/356\\
\hline
 \multicolumn{15}{c}{Table 1. Sample and Fitting Results}
\end{tabular}
 \end{center}
\end{tiny}
\end{sidewaystable}

Observations of GRBs with the {\em Fermi} Gamma-Ray Observatory have revealed additional features beyond the Band function, including a quasi-thermal component\citep{ryde10_zbb,guiriec11_zbb,Axelsson12_zbb,Guiriec15_towards_zbb} and a power-law component extending to high energies\citep{abdo10_zbb_081024b,zhangbb11_zbb}. Through a detailed time-resolved spectral analysis, \cite{zhangbb11_zbb} suggested that the Band component, the quasi-thermal component, and the high-energy component are three elemental spectral components in GRB spectra. A recent detailed study of the time-resolved spectra of three GRBs\citep{Guiriec15_towards_zbb} indeed suggests that the interplay among three elemental components can {explain} a variety of observed GRB spectra. These observations have spurred efforts to apply physical modeling to the GRB data and uncover the physical meaning of various spectral components. The quasi-thermal component is often interpreted as emission from the photosphere of the outflow\citep{meszaros10_zbb_steepslope}. Since this component is predicted to be bright in the standard fireball models, non-detection or a detection of a weak photosphere component point towards a Poynting-flux-dominated flow\citep{zhangpeer09_080916c_zbb} or a hybrid outflow with significant Poynting flux in the jet\citep{gaozhang_15_zbb}. With the quasi-thermal component identified as the photosphere emission, the main Band component is more likely of a synchrotron origin\footnote{Another support to this view is from GRB 110721A, which showed an extremely high $E_p$\citep{Axelsson12_zbb} {far} exceeding the temperature of the photosphere\citep{zhangb12_deathline_zbb,veres_110721a_zbb}.}. \cite{burgess11_zbb,burgess14_zbb} found that the addition of a photospheric blackbody to a synchrotron radiation component reproduces the observed spectrum of eight bright GRBs detected by {\it Fermi}, provided that the electrons are in the {\em slow cooling} regime. This is {however contrary to expected, }since in the GRB emission region, electrons likely cool rapidly due to a strong local magnetic field. The simplest fast-cooling spectra, even with the superposition of the photosphere component, were found incompatible with the observations.

In these and previous studies, the magnetic field in the GRB emission region is assumed to be constant. However, in a rapidly expanding jet source such as a GRB, flux conservation indicates that its magnetic field cannot be preserved as a constant. This important physical consideration has been taken into account by \cite{uhm14_zbb}. By introducing a decrease in the magnetic field with increasing radius as a (conical) jet expands into space and by assuming that the GRB emission region is far from the central engine (typically $10^{15}$ cm and beyond), they create the following picture\citep{uhm14_zbb}: the GRB emission is assumed to be generated from fast-cooling synchrotron radiation of electrons accelerated in the emission region (either internal shocks or internal magnetic reconnection sites), which is relatively far away from the central engine. Electrons are continuously injected as a power law distribution ($N(\gamma) \propto \gamma^{-p}$) with a minimum Lorentz factor $\gamma_{\rm inj}$, and with a certain injection rate as a function of time $R_{\rm inj} = R_{\rm inj}^0 (t/t_0)^q$. Since the emission region is streaming outwards in an expanding jet, the co-moving magnetic field strength decays with radius as $B'(r) = B'_0 (r/r_0)^{-b}$. The accelerated electrons undergo synchrotron cooling in the co-moving frame. Since the cooling time scale is shorter than the dynamical time scale, the electrons are in the fast cooling regime, so that an electron population below $\gamma_{\rm inj}$ is developed as a function of time. In the standard fast cooling model where $B'$ is assumed a constant, the electrons below $\gamma_{\rm inj}$ has a spectrum of $N(\gamma) \propto \gamma^{-\hat p}$ with $\hat p=2$, which gives a photon index $\alpha = -1.5$, inconsistent with the data. However, in a decaying $B'$ field, the degree of cooling varies as a function of time, so that the electron spectrum below $\gamma_{\rm inj}$ is curved with an effective $\hat p < 2$. The photon spectrum therefore hardens. Varying the parameters, the predicted effective $\alpha$ can be as hard as -0.8, with the typical value $\alpha \sim -1$ reproduced for a range of parameters. This concept for the first time introduces a physical fast-cooling synchrotron model that accounts for the typical observed $\alpha$ values\footnote{\cite{zhao14_sync_zbb} proposed a different scenario by invoking decay of magnetic field locally behind the shock front within the framework of the internal shock model. They showed that by considering both synchrotron and inverse Compton cooling of electrons, the typical $\alpha \sim -1$ {can also be reproduced within} that model. With a small emission radius, this model may {struggle to} provide a natural explanation {for} the seconds-long pulses observed in GRBs as well as the hard-to-soft and tracking behaviors of $E_p$ within these pulses.}.

The investigation of this concept has {thus far} been limited to numerical calculations that have produced spectral parameters and temporal evolution properties similar to {those} observed in GRBs\citep{uhm14_zbb}, but the model has {\it not} been applied to real data. In this paper, we aim to test whether such a fast-cooling synchrotron model can really account for the observed GRB spectra. We describe our data reduction method in \S 2 and fitting tools in \S 3. Our fitting results are presented in \S 4, followed by a summary and discussion in \S 5.

\section{Data Reduction }

As will be discussed in \S \ref{sec:lucasfitting}, it is computationally expensive to fit the observed spectra with the numerically calculated synchrotron model. We select one typical GRB detected by {\em Fermi}/GBM as a test, whose time-resolved spectra can be well fit by the Band function. Because most GRBs have strong spectral evolution during the prompt emission phase\citep[e.g.][]{zhangbb11_zbb,lu12_zbb,Guiriec15_towards_zbb}, this work focuses on the time-resolved data. We select the second brightest\footnote{The brightest one is 130427A, which is subject to saturation in the GBM TTE data, and {hence not selected.}} GRB in the {second} GBM catalog, i.e., GRB 130606B (GBM trigger number 392212536/130606497). The brightness of the GRB 130606B enables us to perform {a detailed} time-resolved spectral analysis with fine time-resolution, {owing to the} large number of $\gamma$-ray {photons.} This GRB was about 91 deg from the LAT boresight at the time of the trigger  and remained well outside the nominal field of view for the standard data analysis during the whole prompt emission ($\sim$ 100 s). A LAT detection during the prompt phase was reported using the LAT Low-Energy (LLE) data \citep{Vianello_130606Bgcn_zbb}, but the energy response of the LLE data at this large offset angle to the boresight is not reliable. We thus omit LAT data from our analysis.

\begin{figure}
\includegraphics[scale=0.22,angle=0]{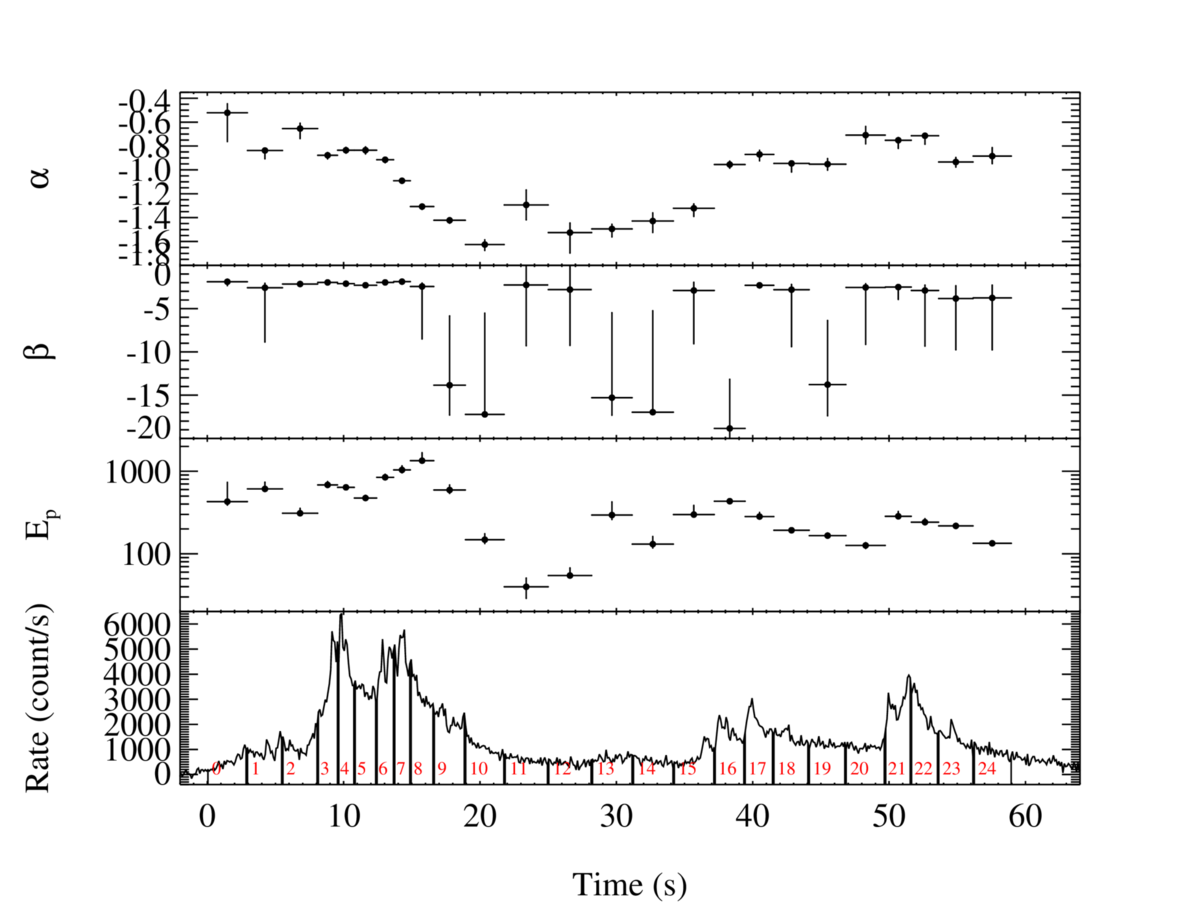}
\caption{{\it Bottom:} Light curve of GRB 130606B in 15-350 keV of {{\em Fermi} GBM detector} NaI 8. Red numbers mark the slice intervals listed in Table 1. {\it Top 1-3 panels}: {Evolution of the Band function parameters via spectral fits of each time interval.} }
\label{fig:bandfit}
\end{figure}

The {\it Fermi}/GBM data are processed using {the standard procedure described in} \cite{zhangbb11_zbb}. We selected the data from the two brightest NaI detectors and the brightest BGO detector. The light curve of the brightest NaI detector (NaI 8) is shown in the bottom panel of Figures \ref{fig:bandfit} and \ref{fig:lucas_fit_evo}. Our analysis only focuses on the main prompt emission episode (from 0 s to 59 s; roughly the episode covered by T$_{90}$) of this burst. Based on the signal-to-noise ratio, we divide this time interval into 24 slices (Table 1 and Figures \ref{fig:bandfit} and \ref{fig:lucas_fit_evo}). For each slice we extract the source spectrum, background spectrum and the corresponding response files, which are needed to perform the spectral fitting. 

\section{A General Monte-Carlo Fitting Approach}

Generally speaking, performing a fit does not only mean finding the ``best-fit" parameters, that bring a particular model into optimum consistency with the data, but also means, more importantly, to find the true likelihood map of the entire parameter space, with which one can know how the model can be really constrained in a global confidence manner. The latter part is more challenging since theoretical models always involve multiple model parameters that may be subject to correlation and multi-modality of the likelihood in their parameter space. Traditional fitting algorithms (e.g., Levenberg-Marquardt) typically fail to map the multi-dimensional likelihood. {We have} developed a fitting engine ({\it McFit}) that employs a Bayesian Monte-Carlo (MC) fitting technique to reliably fit parameters that are constrained by the data even {when} other parameters are unconstrained. Such a technique has been successfully applied to {\it Swift} afterglow modeling of GRBs with 7 parameters\citep{zhangbb14_zbb}. Using this technique, the best-fit parameters and their uncertainties can be realistically determined by the converged MC chains. 

{To compare a theoretical spectral model to data, we need to convolve the model with instrumental {response, the Detector Response Matrix (DRM).} We have written a general-purpose forward-folding code that can handle the instrument response $R(I,E)$ from any mission (e.g. {\it Fermi}/GBM), read in any model spectrum, $F(E, P)$, regardless of its form (e.g. a mathematical function or a tabulated array), and calculate the model-predicted count spectrum $C_M(I,P)=\int_{0}^{\infty}F(E,P)R(I,E)dE$, which can be directly compared with the observed count spectrum $C(I)$.} We then calculate the likelihood\footnote{In this paper, we use the maximum likelihood-based statistics, the socalled PGSTAT, for Poisson data, given by Cash (1979), but assuming the background to be Gaussian; see also \url{https://heasarc.gsfc.nasa.gov/xanadu/xspec/}} for those $C_M(I, P)$ and $C(I)$ pairs. Such a likelihood function is then input into our fitting engine {\it McFit} so that the parameter space can be explored. A final software package, {\it McSpecFit}, {combines} this Bayesian MC engine, the general forward-folding codes and the likelihood calculations, and {can} fit any models to the observed spectra.

\section{Results}

\subsection{Band Function Fits}

{We first} fit the observed time-resolved spectra with the empirical Band function\citep{band93_zbb}:
\begin{equation}
\begin{tiny}
\label{eqn:band}
F(E) = A\left\{
\begin{array}{ll}
	\left(\frac{E}{100\text{ keV}}\right)^\alpha e^ {\left[-\frac{E}{E_\text{0}}\right]},
 & E\leq (\alpha-\beta)E_\text{0}, \\
	\left(\frac{E}{100\text{ keV}}\right)^\beta e^{\left(\beta-\alpha\right)} \left(\frac{(\alpha-\beta)E_\text{0}}{100\text{ keV}}\right)^{\alpha-\beta},
& E> (\alpha-\beta)E_\text{0},
\end{array}
\right.
\end{tiny}
\end{equation}
where
\begin{equation}
\label{eqn:Ec}
E_\text{p}=(2+\alpha) E_0,
\end{equation}
$A$ is the normalization factor at 100~keV in units of photons~s$^{-1}$~cm$^{-2}$~keV$^{-1}$, $\alpha$ is the low-energy photon spectral index, $\beta$ is the high-energy photon spectral index, and $E_\text{p}$ is the peak energy in units of keV in the observed $\nu F_\nu$ spectrum.

The results of our fits for each slice are shown in Table 1. The measurement of the goodness-of-fit in each slice is plausible, with PGSTAT/dof $\sim$ 1, as shown in the 5th column of Table 1. As an example, we show the parameter constraints resulting from our {\em McSpecFit} in the inner plot of Figure \ref{fig:contour}. We also plot the evolution of the three characteristic parameters of the Band function model in Figure 1. We notice that except for 9 slices (intervals 0, 2, 3, 4, 5, 17 and 21) the values of $\beta$ are constrained to have fairly large values (2.5 -- 8) and large uncertainties, which indicates that a cutoff power law model would also give equally good fits. We therefore mainly focus on the properties of $\alpha$ and $E_p$.

{A strong spectral evolution is shown in Figure 1. $E_p$ and the peak count rate (or equivalently, peak flux) in each slice correlate with each other and show a clear tracking behavior (see also Ford et al 2015, Lu et al 2012).}

\subsection{Synchroton Model Fits}
\label{sec:lucasfitting}

Next, we apply the fast-cooling synchrotron emission model with a decaying magnetic field\citep{uhm14_zbb} to fit the time-resolved spectra.

Mathematically, the \cite{uhm14_zbb} model is defined by the following {\it nine} parameters: 
\begin{itemize}
\item the bulk Lorentz factor $\Gamma$ of the emitting region;
\item the magnetic field strength $B_0$ {(in units of {\it Gauss})} at $10^{15}$ cm;
\item the magnetic field decay index $b$;
\item the power-law index $p$ of the injected electron spectrum;
\item the minimum injection Lorentz factor $\gamma_{\rm inj}$ of electrons;
\item the power-law index $q$ of the injection rate as a function of time;
\item the {normalization} of the electron injection rate $R_{\rm inj}^0$ {in units of s$^{-1}$} ;
\item the observer-frame time since the first electron injection $\hat t$ {in units of s} of a given emission episode;
\item the redshift $z$.
\end{itemize}

The theoretical flux density (in unit of mJy) can be calculated numerically in the form
\begin{equation}
F_\nu(E)=M(E; \Gamma, p, \gamma_{\rm inj},R_{\rm inj}^0 , q , B_0 ,b ,\hat t, z).
\label{equ:model}
\end{equation}
We implement the numerical code of \cite{uhm14_zbb} into the data fitting code, having it serve as a black box to generate numerical model spectra given any set of the input parameters.

\begin{figure}
\includegraphics[scale=0.39,angle=0]{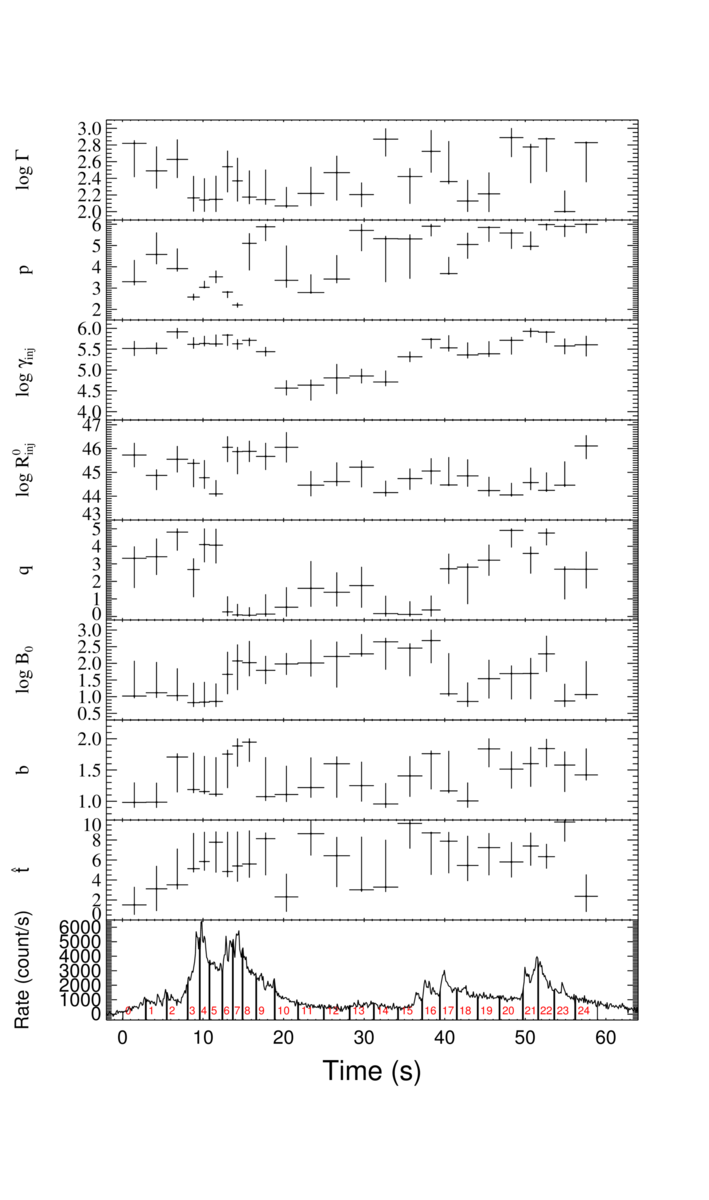}
\caption{{\it Bottom:} Same as the bottom panel of Figure 1. {\it Top 1-8 panels}: Parameter evolution of the synchrotron model fitting of each slice. }
\label{fig:lucas_fit_evo}
\end{figure}

\begin{figure*} \centering
\begin{tikzpicture}[
 every node/.style={anchor=south west,inner sep=0pt},
 x=1mm, y=1mm,
 ] 
 \node (fig1) at (0,0)
 {\includegraphics[width=7.0in]{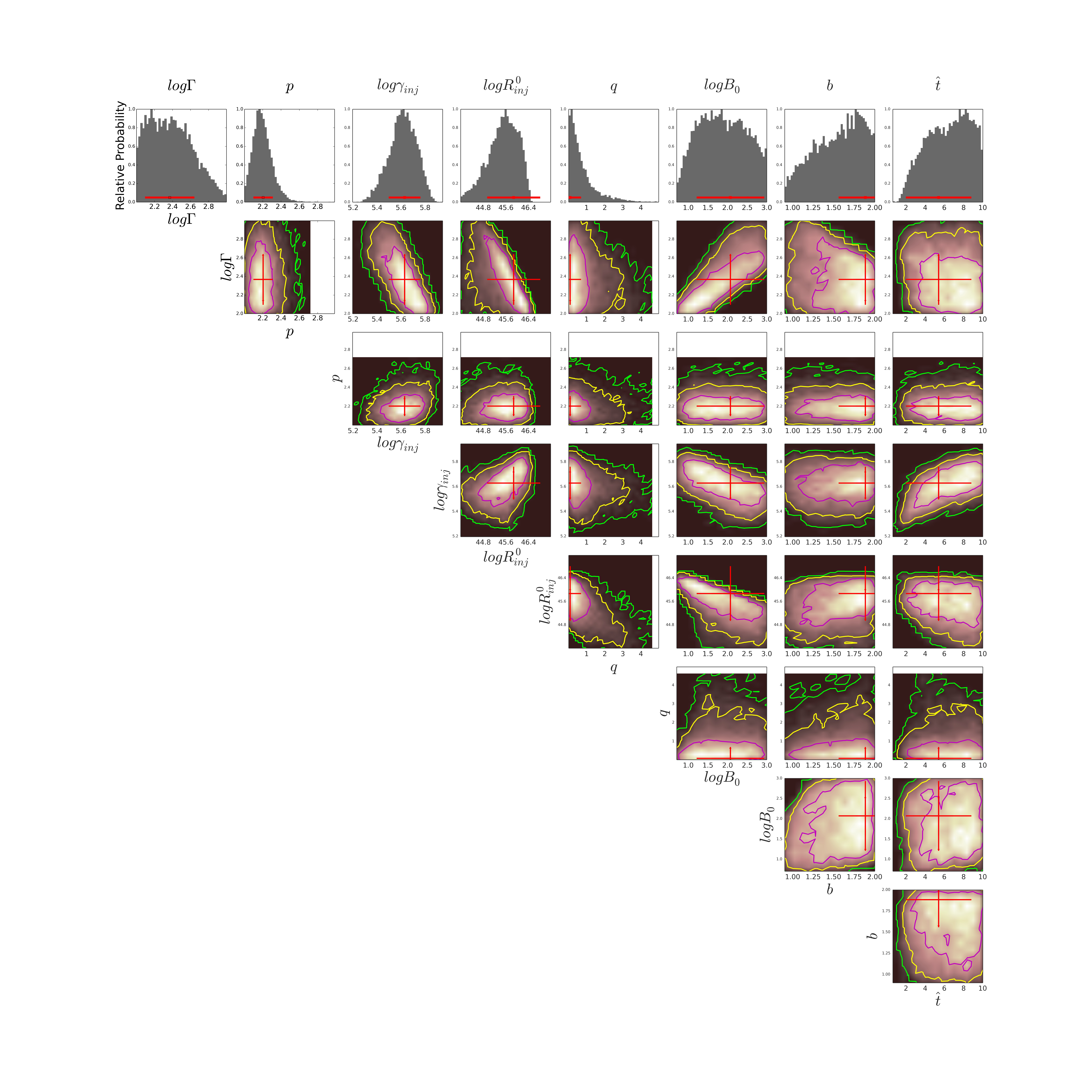}};
\node (fig2) at (15,10)
 {\includegraphics[width=2.7in]{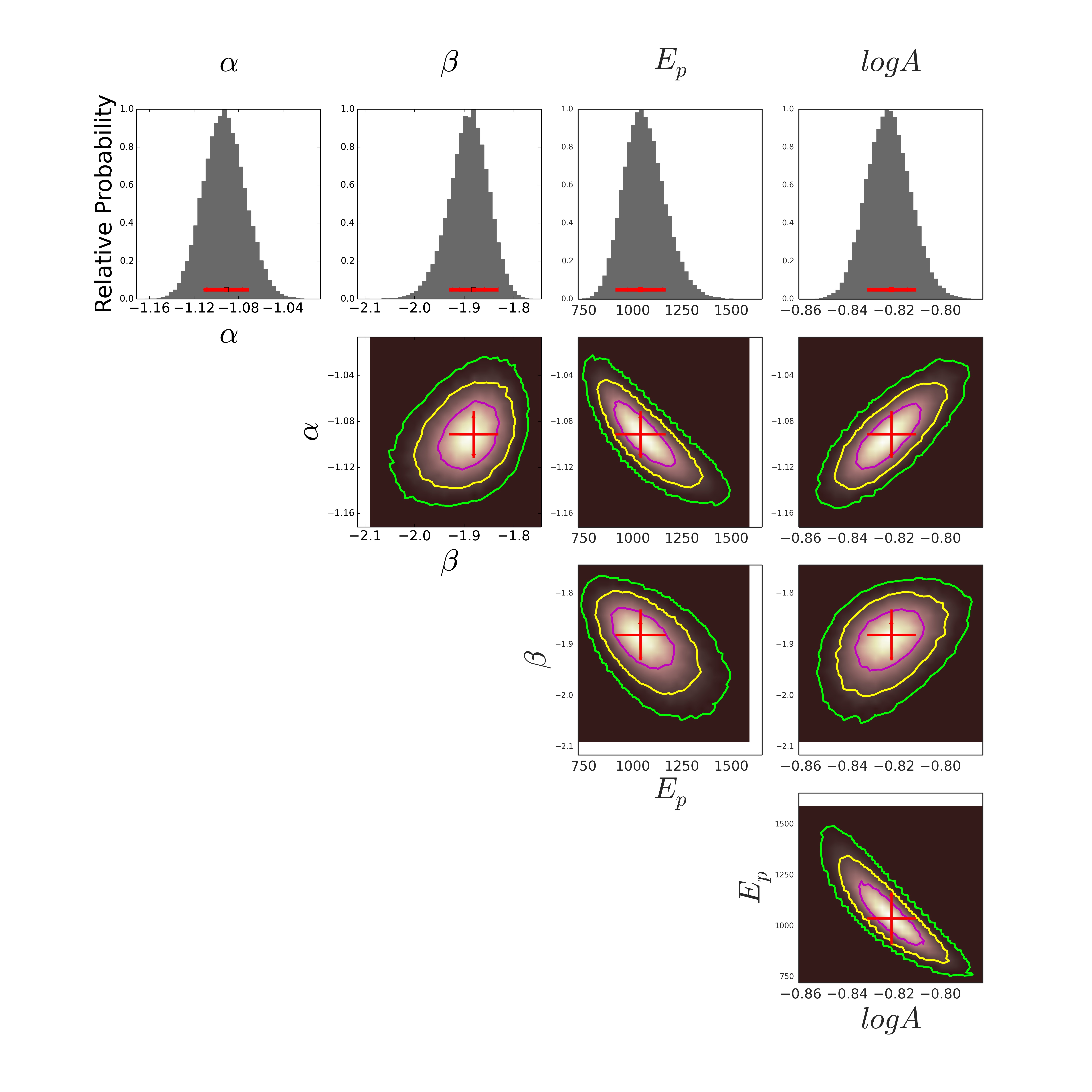}};
\end{tikzpicture}
 \caption{Parameter constraints of the synchrotron model ({\it up-right}) and Band function ({\it bottom left}) fitting for Interval 7. Histograms and contours show the likelihood map of the parameter-constraint outputs by our {\it McSpecFit} package. Red crosses mark the best-fit values and 1-sigma error bars. All parameters are constrained in reasonable
ranges (see Table 1). }
\label{fig:contour}
\end{figure*}

\begin{figure}
\includegraphics[scale=0.2]{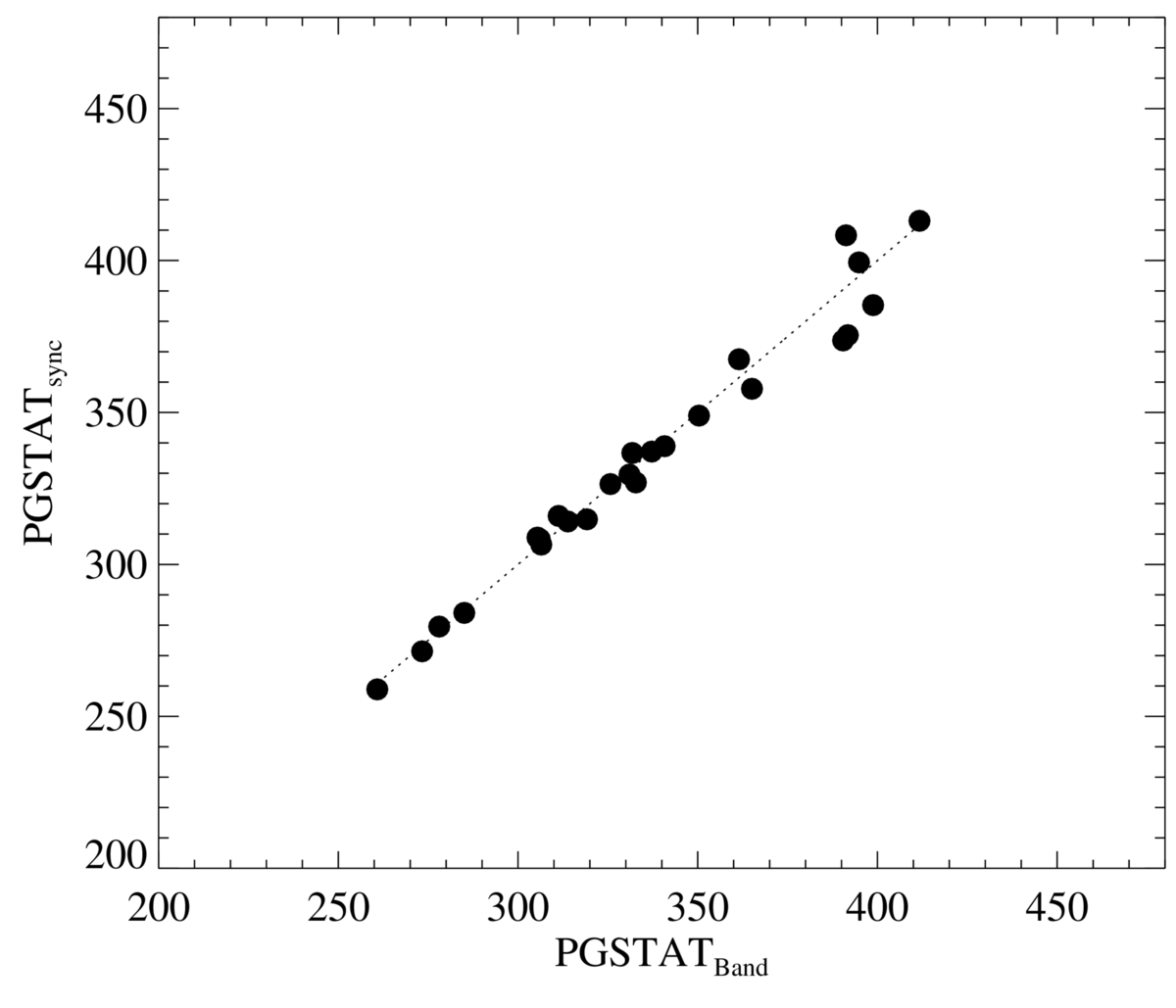}
\caption{Comparison between the PGSTAT of the Band function fitting and the synchrotron model fitting for the 25 time slices in Figure 1. The dashes mark the line of equality. }
\label{fig:comp_pgstat}
\end{figure}

 \begin{figure*} \centering
 \centering
 \begin{tabular}{cc}
 \includegraphics[scale=0.21]{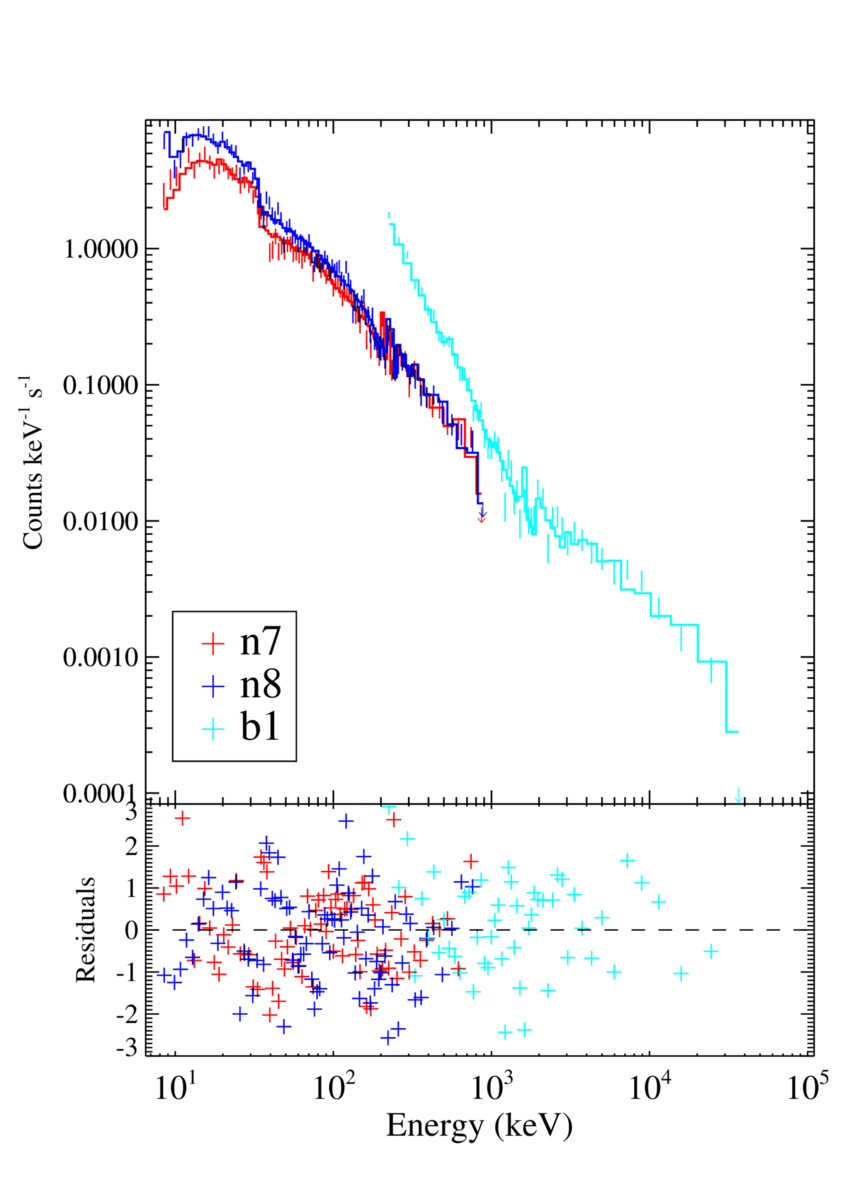} & 
 \includegraphics[scale=0.21]{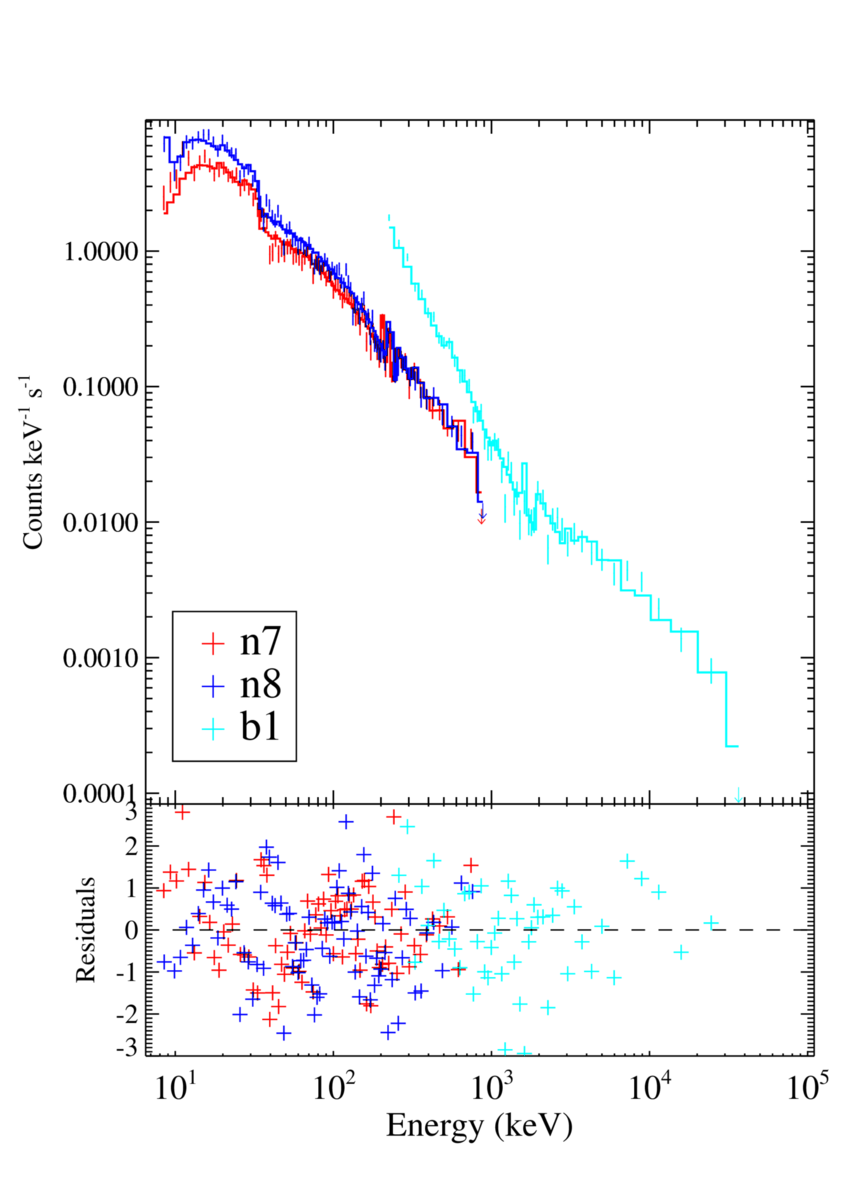} \\
 \includegraphics[scale=0.21]{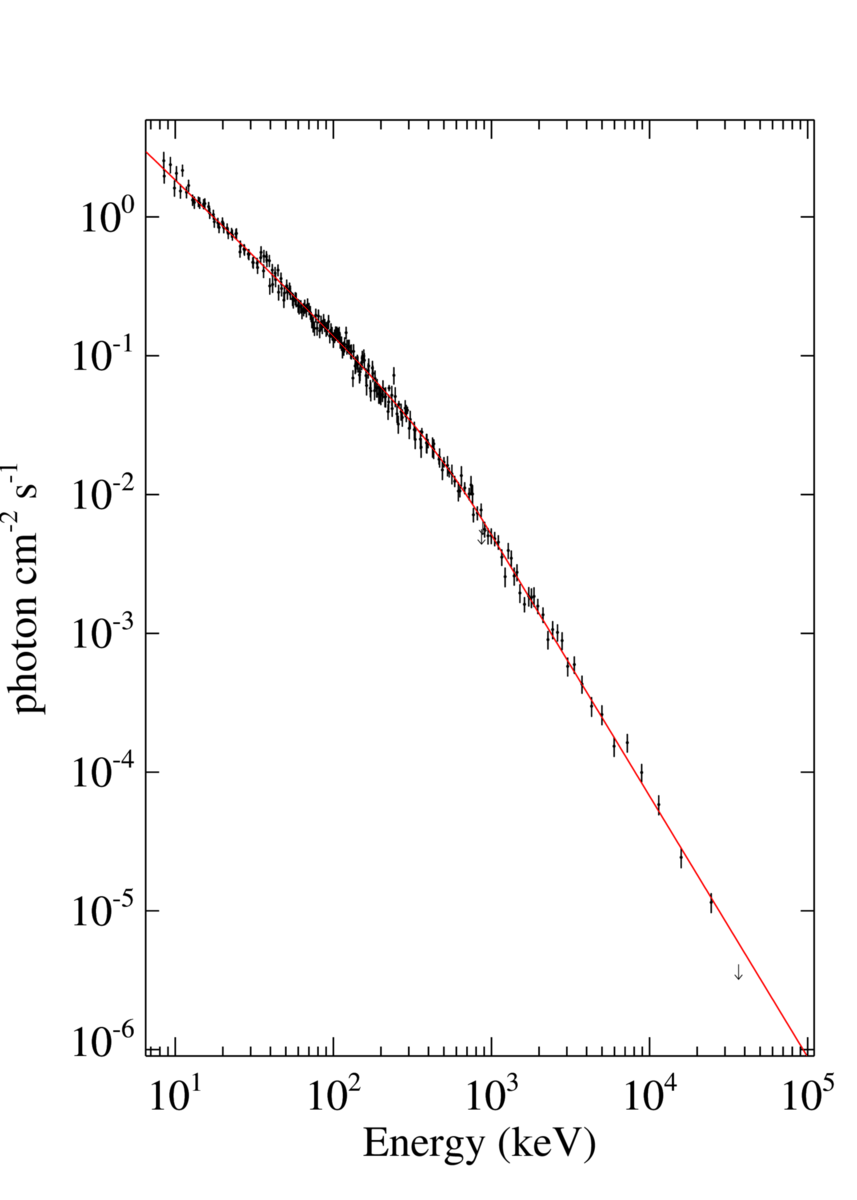} & 
 \includegraphics[scale=0.21]{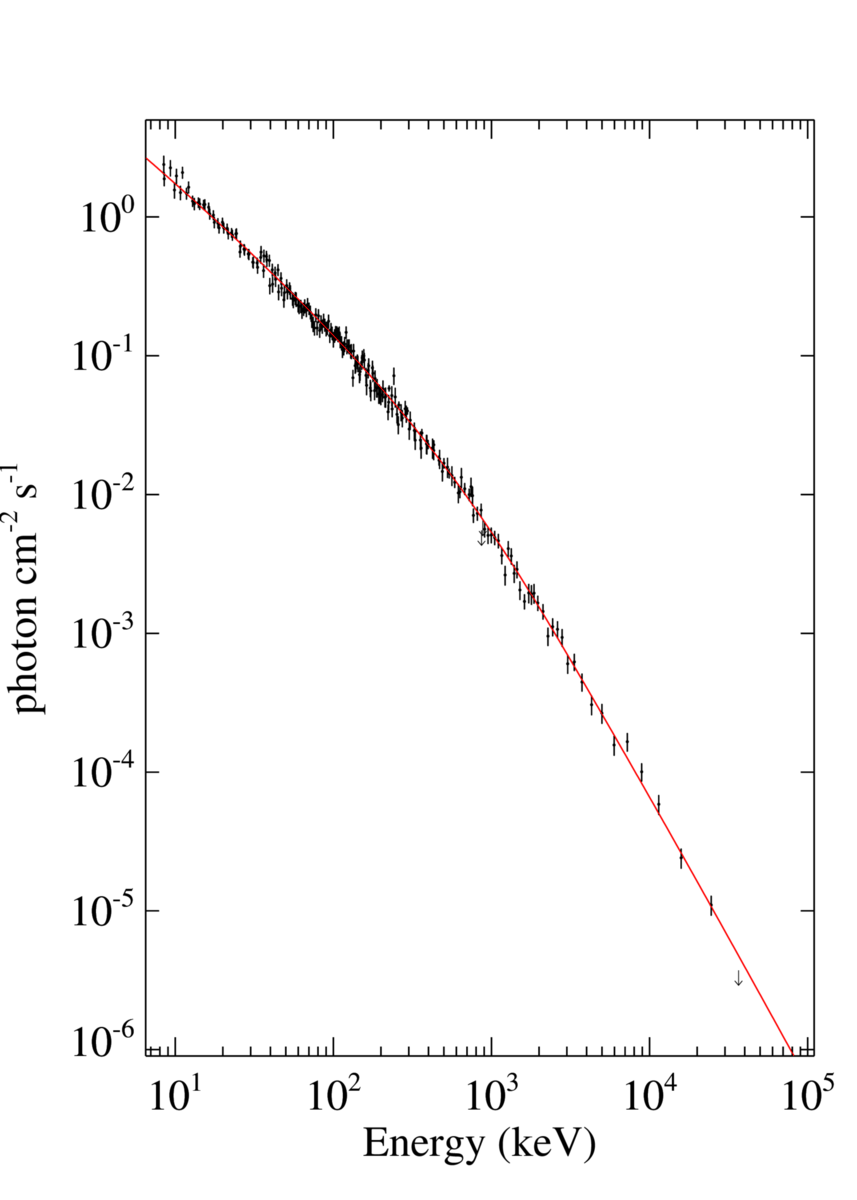} \\

 \end{tabular}
 \caption{Comparison between the Band function fitting and synchrotron model fitting for Interval 7. The spectra have been re-binned into wider bins for clarity. {\it Top two:} observed count spectrum vs. model. {\it Top Left} shows the Band function fitting and {\it Top Right} shows synchrotron model fitting. {Bottom two:} the de-convolved photon spectrum plots. {\it Bottom Left} shows Band function fitting, red line is the theoretical photon spectrum of the Band function and data points are the ``observed" photon flux which is obtained by de-convolving the observed count spectrum using instrument responses. {\it Bottom Right:} same as {\it Bottom Left} but for the synchrotron model fitting where the red line is the theoretical photon spectrum of the synchrotron model. }
\label{fig:ph_obs_comp}
 \end{figure*}

\begin{figure} \centering
\begin{tikzpicture}[
 every node/.style={anchor=south west,inner sep=0pt},
 x=1mm, y=1mm,
 ] 
 \node (fig1) at (0,0)
 {\includegraphics[scale=0.2]{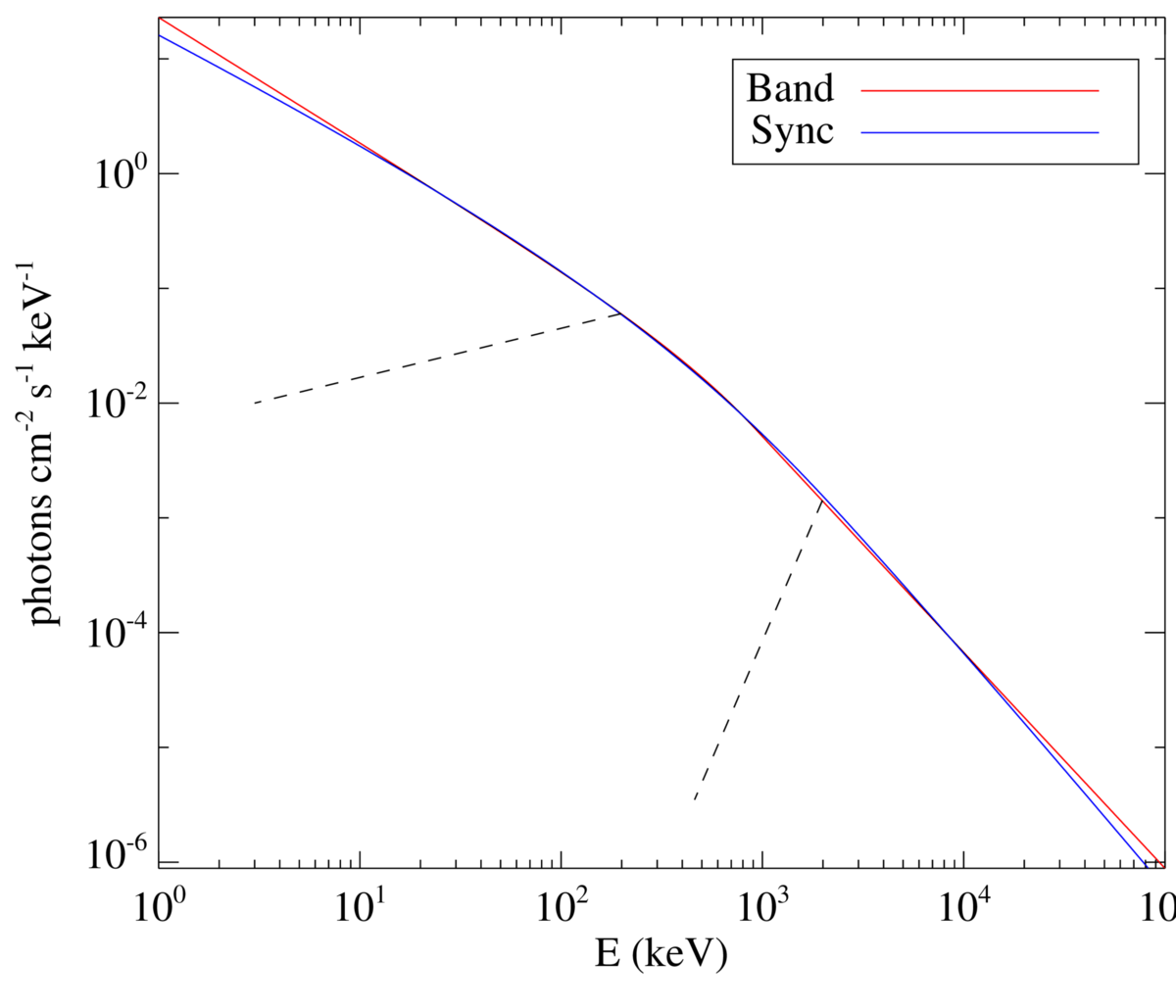}};
 \node (fig2) at (12,11)
 {\includegraphics[scale=0.090]{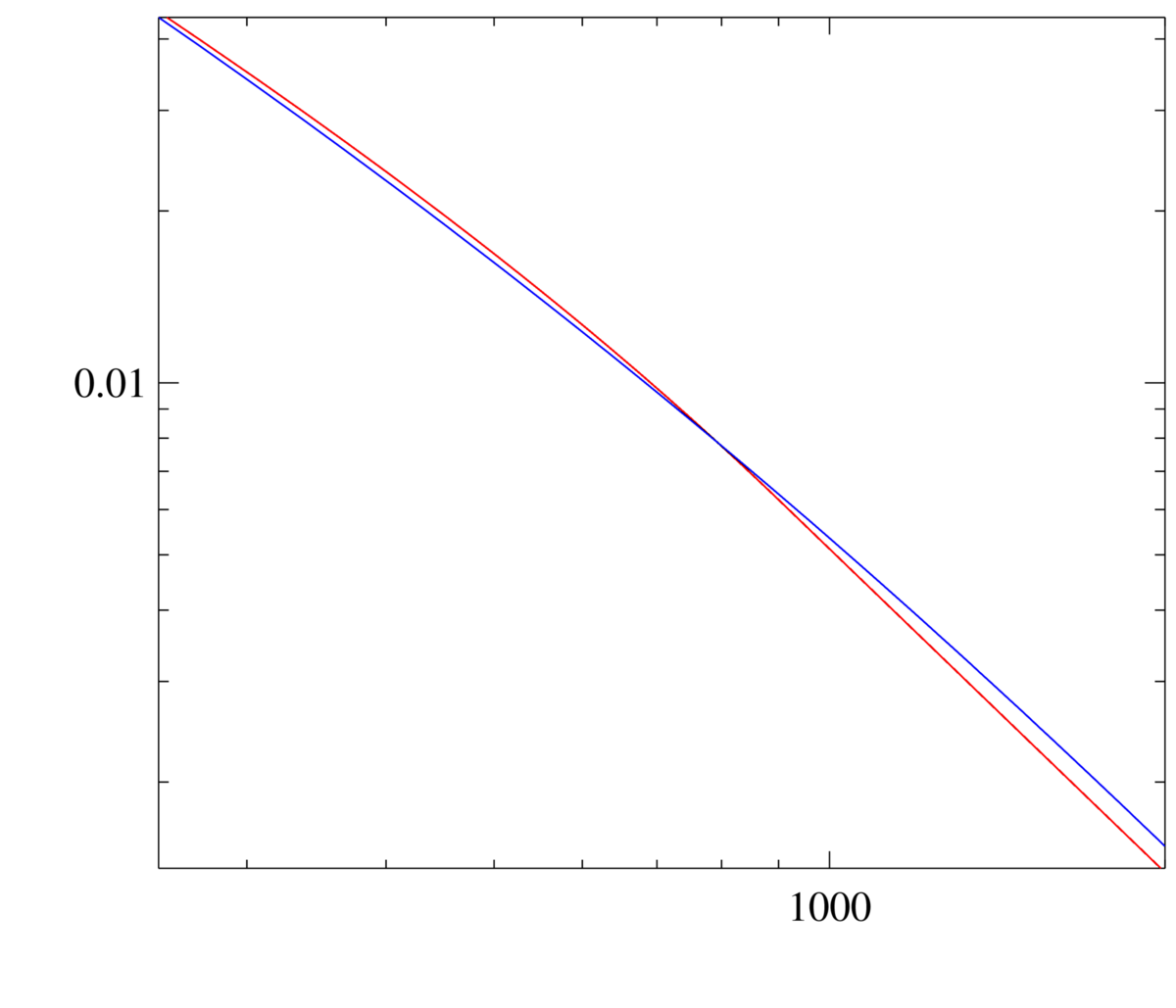}};
\end{tikzpicture}
 \caption{Comparison between the two theoretical photon models in Figure \ref{fig:ph_obs_comp}. }
\label{fig:ph_th_comp}
\end{figure}

The direct output of the modeling is $F_\nu(E)$. It is converted to photon flux and then read into {\em McSpecFit}. Spectral fitting is then performed following the Bayesian Monte Carlo fitting method {within the allowed parameter ranges (Table 2)} as introduced in Section 3, until the best-fit model parameters are obtained. 

\setcounter{table}{1}
\begin{table}

\setlength{\tabcolsep}{2pt}

\caption{Allowed Range of The Parameters of the Synchrotron Model in the Monte Carlo fit}
\begin{center}

\begin{tabular}{l|l}
\hline
Parameter & Range \\
\hline
$\Gamma$ & [$10^2$, $10^3$] \\
$p$ & [2, 6] \\
 $\gamma_{\rm inj}$ & [$10^{3}$, $10^{6}$]\\
$R_{\rm inj}^0$ ($s^{-1}$) & [10$^{44}$, 10$^{50}$]\\
$q$ & [0.01, 5] \\
$B_0$ (G)& [5, 1000] \\
$b$ & [0.9, 2.0] \\
$\hat t$ (s) & [0.1, 10.0]\\
\hline

\hline
\end{tabular}

\end{center}

 \label{tab:range}
\end{table}

 \begin{figure*} \centering
 \centering
 \begin{tabular}{ccc}
 \includegraphics[scale=0.13]{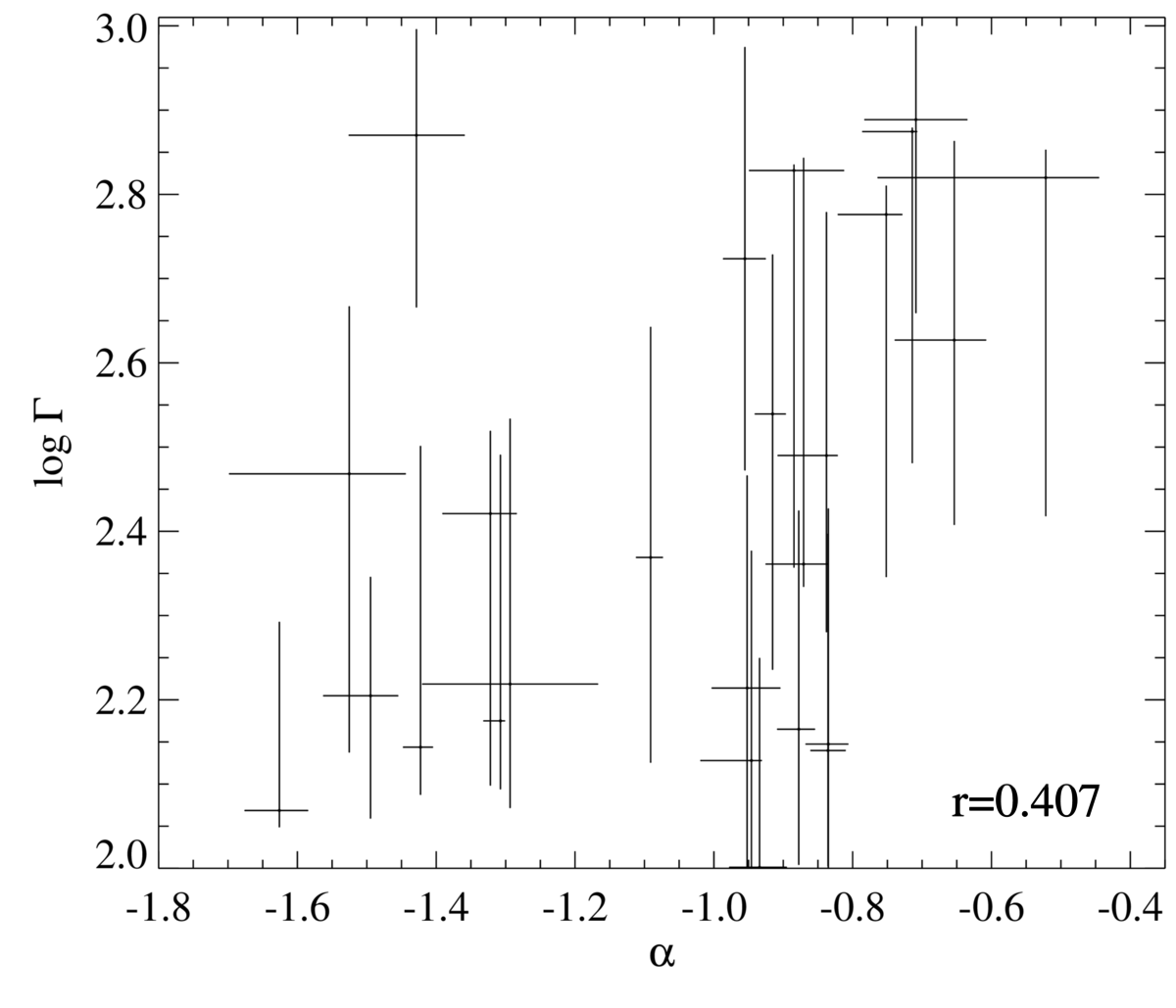} & 
 \includegraphics[scale=0.13]{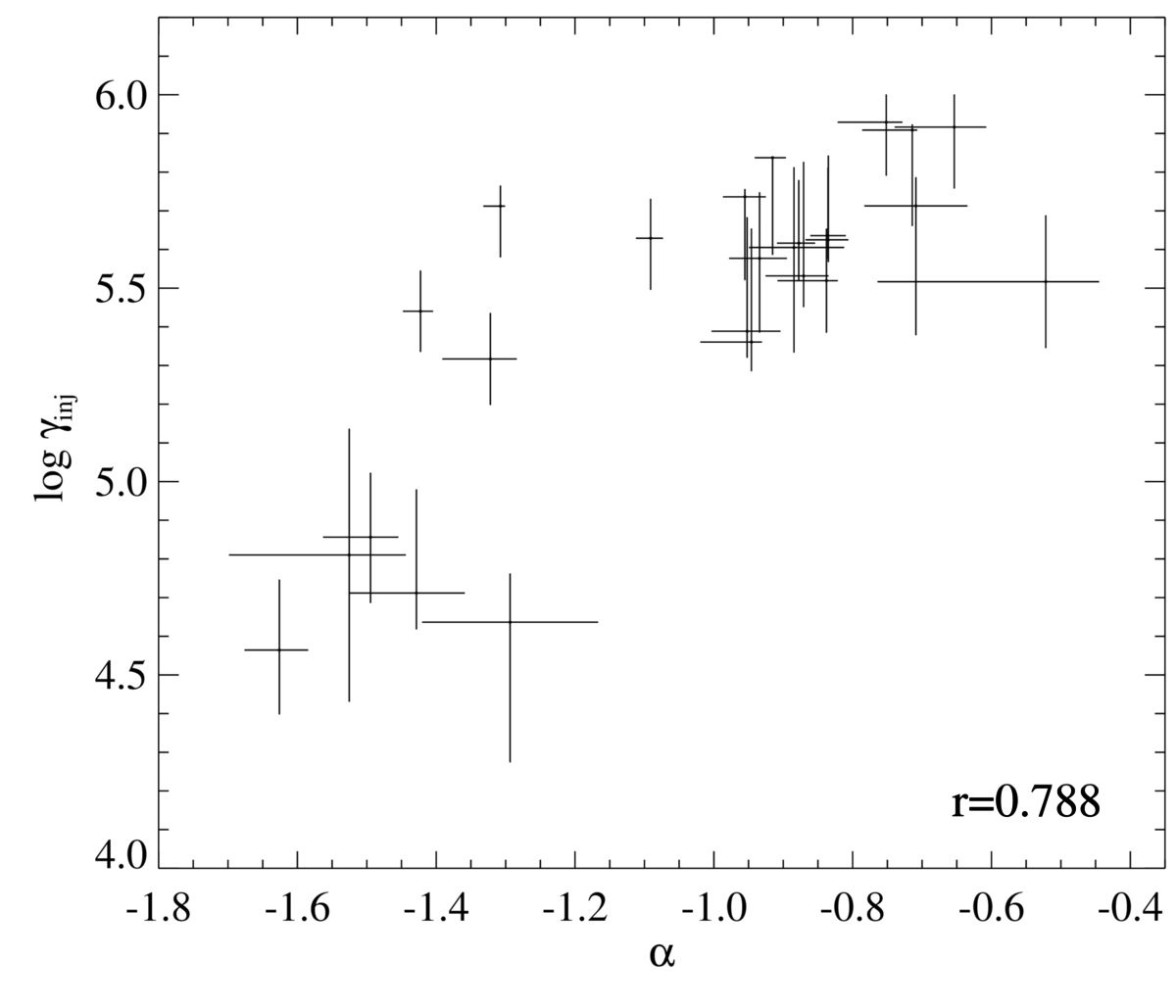} &
 \includegraphics[scale=0.13]{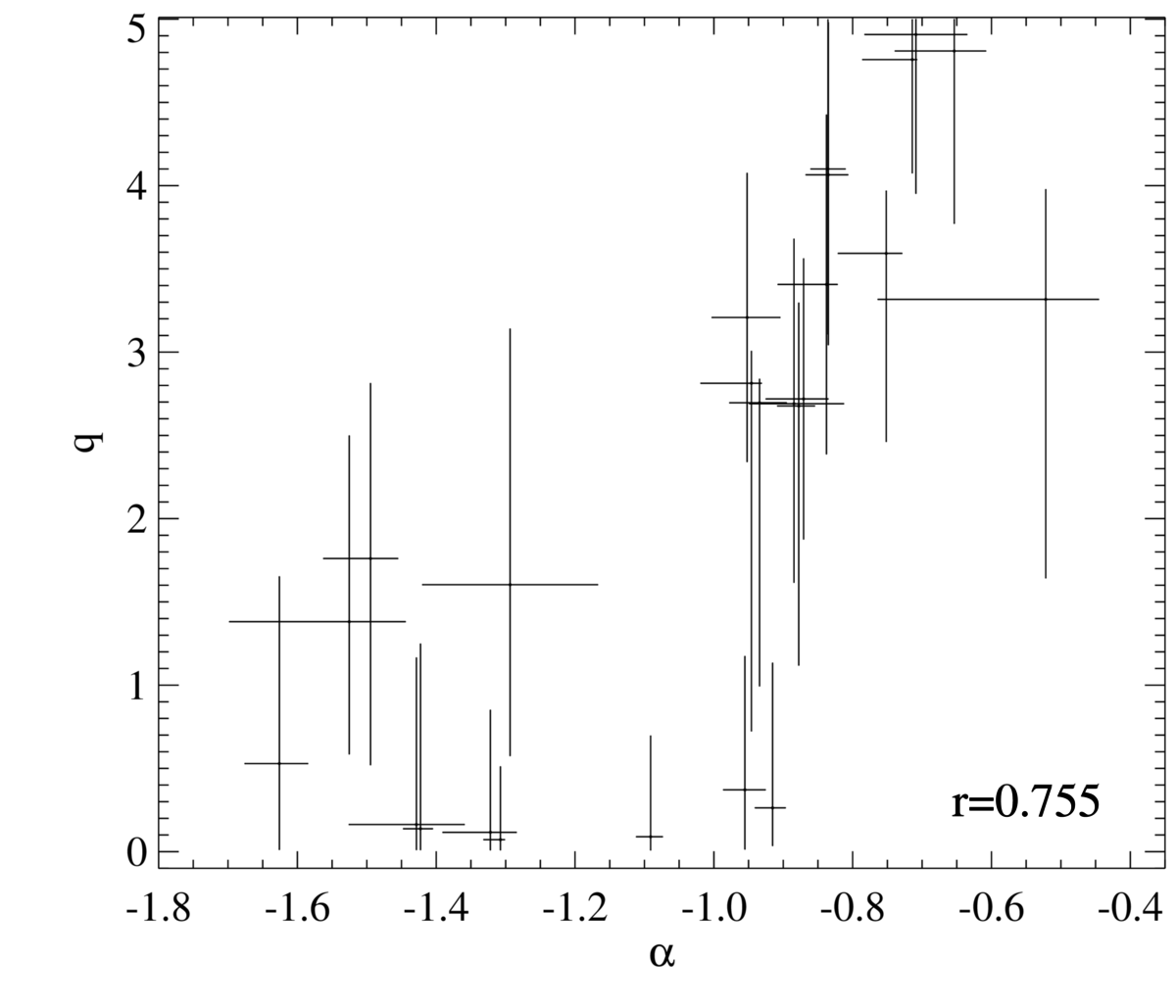}\\ 
 \includegraphics[scale=0.13]{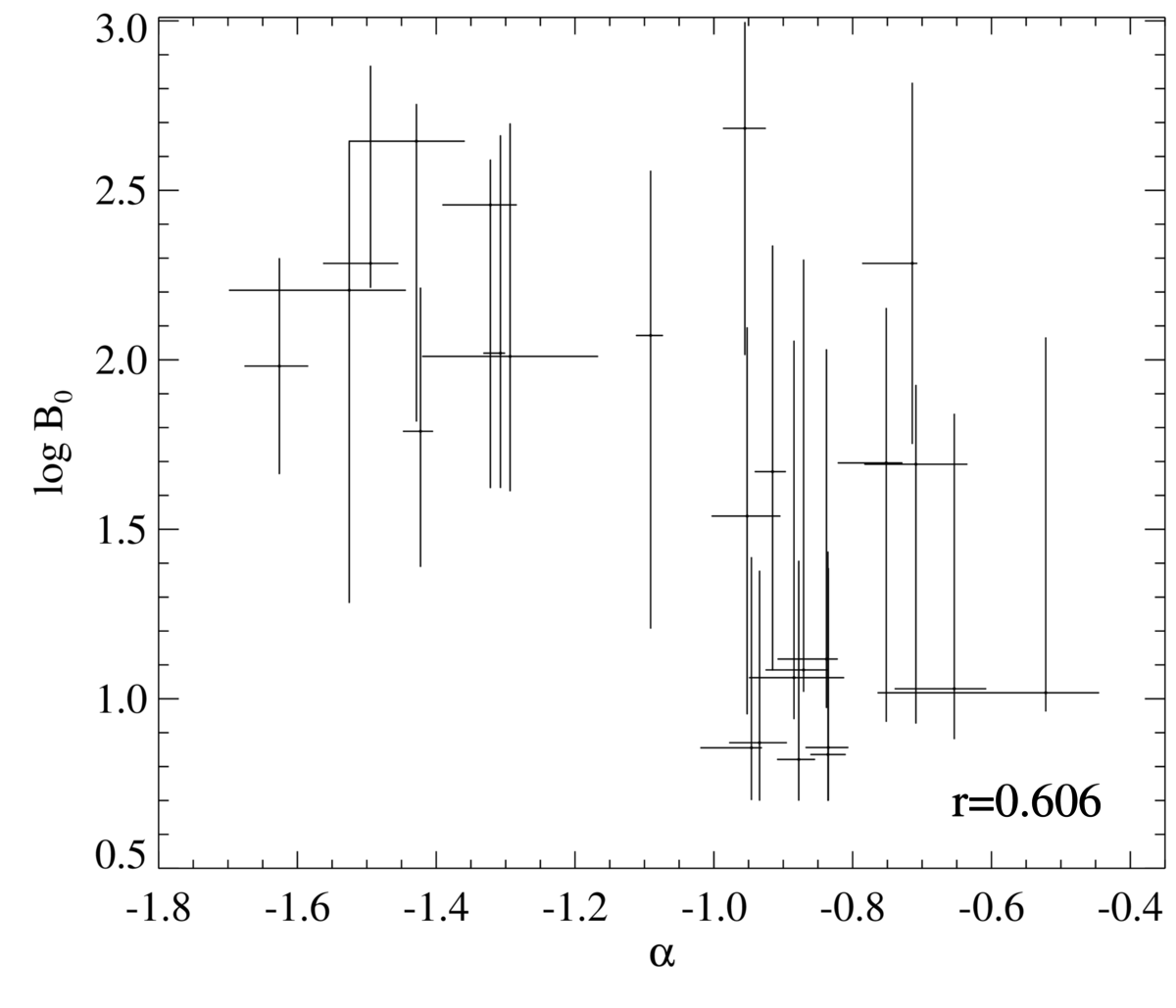} &
 \includegraphics[scale=0.13]{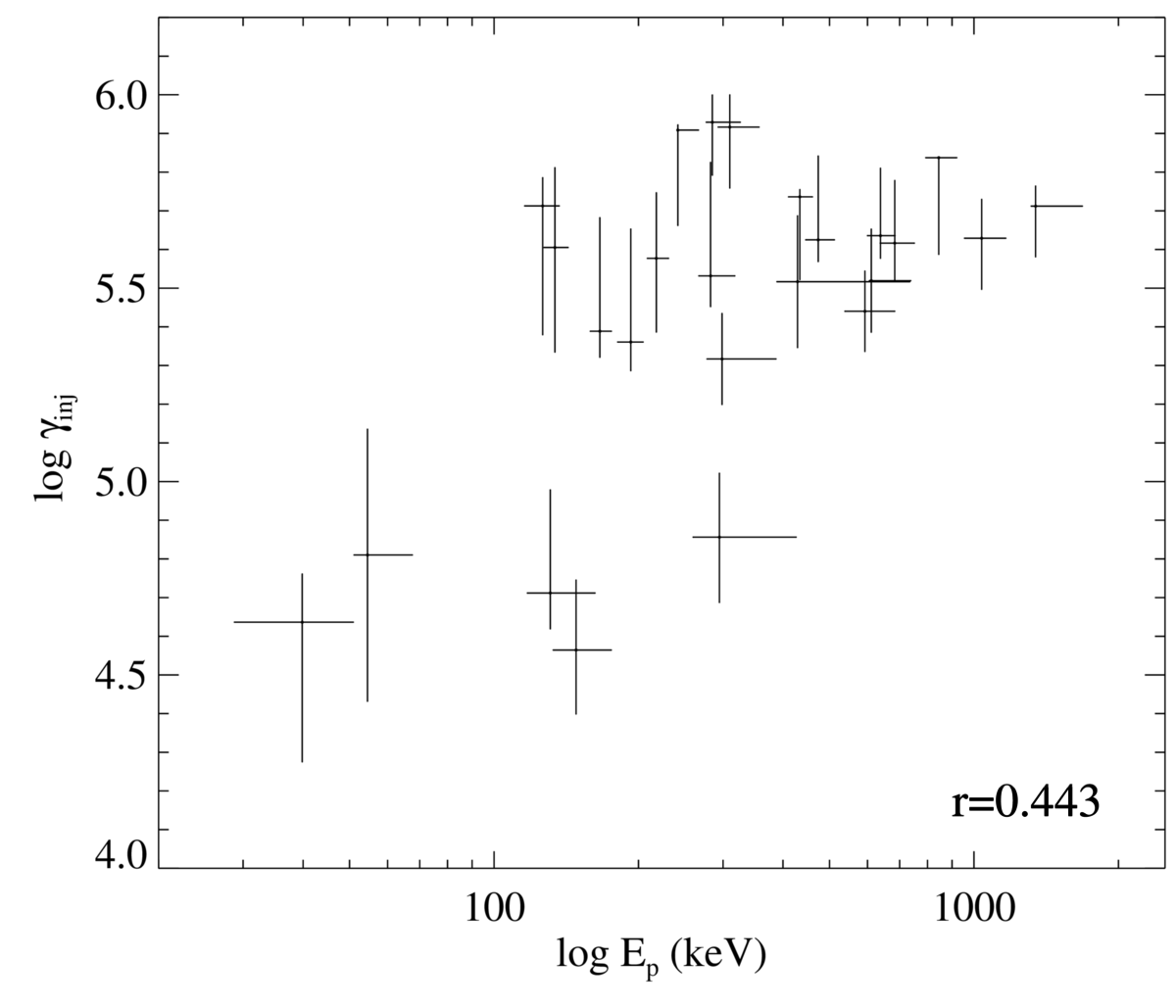} &
 \includegraphics[scale=0.13]{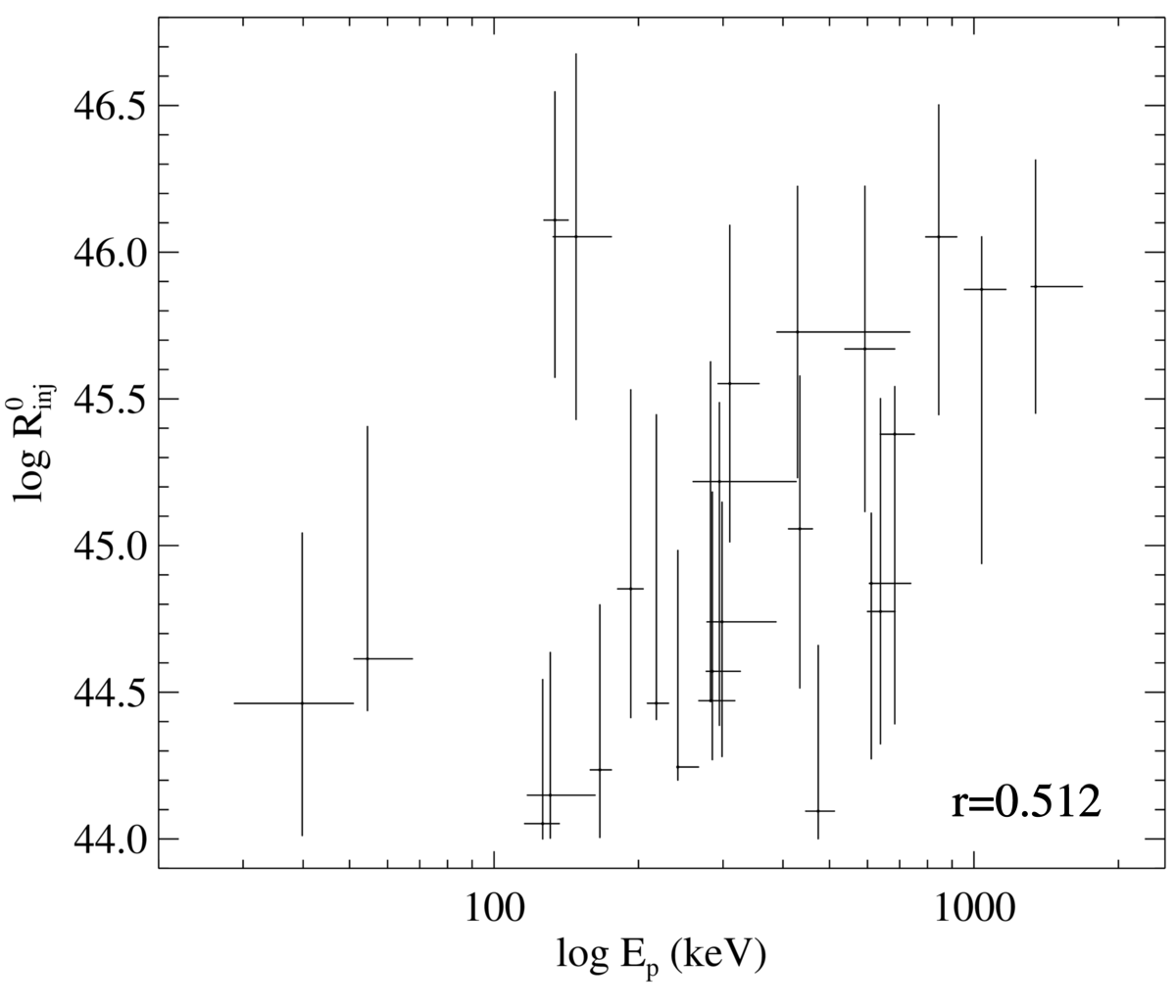}

 \end{tabular}
 \caption{We compare the parameters of the physical model with those of the Band function and show the correlations between those pairs with strong and moderate correlations. The Pearson correlation coefficients of each pair are also marked in each plot. }
\label{fig:corr}

 \end{figure*}

Two caveats are worth mentioning. First, GRB 130606B does not have a measured redshift. We assume $z=1$ in the modeling. The shape of spectrum does not depend on the redshift. For a different redshift, the same observed spectrum can be reproduced by essentially adjusting the injection rate parameter $R_{\rm inj}^0$. Second, in the fitting, $\hat t$ is taken as a free parameter. Strictly speaking, it should be defined as the observer-frame time since the injection of the first group of electrons in a particular emission episode. However, since GRB light curves include {overlapped pulses}, it is hard to track the number of emission episodes. Furthermore, the predicted flux is not very sensitive to $\hat t$ if the duration is in the time scale of seconds\citep{uhm14_zbb}. This can be readily seen in Fig.\ref{fig:contour}, which shows that the contour of $\hat t$ is relatively broad. We therefore make no effort to guess the $\hat t$ value for each time bin, and leave it as a free parameter.

The results of fitting the data to the synchrotron model are presented in Table 1 and Figure \ref{fig:lucas_fit_evo}. Similar to the Band function fitting, we show an example of the parameter constraints for the same time interval 7 in the outer plot of Figure \ref{fig:contour}. We show that the synchrotron model {provides} equally good fits to the data as the empirical Band function model (Figure \ref{fig:comp_pgstat}). 

The evolution of various model parameters is shown in Figure \ref{fig:lucas_fit_evo}. Similar to the Band function parameters, these model parameters all undergo variations throughout the burst. As expected, the magnetic field strength in the emission region is found to decay with radius, with $b$ varying in the range between 1 and 2. However, the large error of $b$ makes it possible that $b$ does not evolve significantly throughout the burst. This is consistent with the expectation: whereas the flux and $E_p$ vary significantly throughout the burst due to the change of electron injection rate and typical Lorentz factors, the geometric configuration {and the degree of magnetic dissipation} of the jet (which defines $b$) do not evolve significantly.

\subsection{Comparison between the Two Model Fits}

In Figure \ref{fig:comp_pgstat} we compare the goodness of fits, PGSTAT, of the Band model and the synchrotron model. One can see that the two models give essentially identical goodness-of-fit. This shows that the {\it physical} model proposed by \cite{uhm14_zbb} successfully fits the typical GRB spectra which are otherwise fitted by the empirical Band function (virtually a mathematical shape). In Fig.\ref{fig:ph_obs_comp}, we show a comparison between the photon spectra of the two model fits as well as the residuals of the fits for the time interval 7 as an example. Again it clearly shows that both models are adequate to represent the data. In Fig.\ref{fig:ph_th_comp}, we show the model curves of the Band function and synchrotron model. One can see that the two curves are essentially identical within the GBM energy band (see also \citealt{uhm14_zbb}), and only show noticeable difference in a much wider energy band. This explains the capability of the synchrotron model to mimic the Band function within the GBM band.

The $E_p$ tracking flux feature (Figure \ref{fig:bandfit}) shown in this burst is commonly seen in a large fraction of GRBs\citep[e.g.][]{ford95_zbb,lu12_zbb}. For the simplest synchrotron model with a constant $\gamma_{\rm inj}$, a hard-to-soft evolution pattern is predicted\citep{uhm14_zbb}. Our analysis suggests that by allowing variation of the parameters especially $\gamma_{\rm inj}$, the synchrotron model can also reproduce $E_p$-tracking patterns in GRBs.

The hardest $\alpha$ in our Band function fit is $-0.52^{+0.08}_{-0.24}$ (interval 0). The best-fit value -0.52 already exceeds the so-called synchrotron ``line-of-death''\citep{preece98_deathline_zbb}. However, due the large error of $\alpha$, the data are also consistent with the synchrotron model. This suggests that ``line-of-death'' is not a hard limit for the synchrotron model.

\subsection{What makes the Band Spectrum?}

{A relevant question is what physical processes shape a synchrotron spectrum to be ``Band-like". In order to address this point, we systematically checked the pair correlations between the fitted Band parameters and the synchrotron model parameters. Among those correlations, we find the following two pairs (Figure 7) showing ``strong'' (Dancey \& Reidy 2004) correlations with Pearson correlation coefficients, $|r|>0.7$: $\alpha$-$\gamma_{\rm inj}$ and $\alpha$-$q$; and the following four pairs showing ``moderate" correlations with $0.4\le |r| \le 0.6$: $\alpha$-$\Gamma$, $\alpha$-$B_0$, $E_p$- $\gamma_{\rm inj}$, $E_p$-$R_{inj}^0$. This indicates that the observed spectral properties can be mostly accounted for by the variations of the jet magnetization and the electron injection parameters.

There are two physical ingredients entering the modeling:
\begin{itemize}
\item The first is the cooling of electrons and their synchrotron radiation, which is well understood \citep{uhm14_zbb}. This physical ingredient controls the {\em shape} of the spectrum. The general Band-like spectral-shape can be reproduced regardless of the details of the model parameters. 
\item The second physical ingredient is related to the energy dissipation physics and particle acceleration physics, which controls the model parameters such as $B_0$, $b$, $q$, $\gamma_{\rm inj}$, and determine the Band parameters $\alpha$, $E_p$, and $\beta$. This likely invokes magnetic reconnection in a highly variable, highly magnetized environment, and the detailed physics is messy and poorly studied\footnote{Recent numerical simulations \citep[e.g.][]{deng15_zbb} are starting to tackle these difficult problems.}.
\end{itemize}
 As a result, one cannot predict from the first principles the values of those parameters, but can only infer them from the data. In any case, the inferred parameters as listed in Table 1 fall into the reasonable ranges from expectation\citep{uhm14_zbb}, and the apparent correlations among parameters are also generally understandable. }

\subsection{Non-existence of the Thermal Component}

For GRB 130606B, the Band function or the synchrotron model can adequately fit the data. There is no need to introduce a quasi-thermal component, which is presumably the emission from the photosphere of the outflow. The non-detection of the photosphere component poses a significant constraint on the GRB jet composition, which should have a significant fraction of Poynting flux in the outflow\citep{zhangpeer09_080916c_zbb,gaozhang_15_zbb}.

\section{Summary and Discussion}

In this paper, we successfully fit the time-resolved spectra of GRB 130606B with a fast-cooling synchrotron radiation model that incorporates magnetic field decay. The same data are adequately fit by the empirical Band function. This is the first time that we give Band function a physical meaning through directly fitting the real GRB data with a {\it physical} model {incorporating fast-cooling synchrotron radiation}. 

Our fast-cooling synchrotron model\citep{uhm14_zbb} involves a minimal number of physical assumptions. It is impressive that the time-resolved spectra of GRB 130606B, which are characterized by the Band function with a wide range of spectral parameters, all can be equally well fit by the synchrotron model. An $\alpha$-index as hard as $-0.52^{+0.08}_{-0.24}$ can be reproduced by the synchrotron model. 

Limited by the computational resources, it is beyond the scope of this paper to fit the spectra of a large sample of GRBs. With a 3.5 GHz 6-Core Intel Xeon E5 CPU on a Mac Pro workstation, it took about 160 CPU hours to finish synchrotron model fit for one time slice. Altogether it took about 4000 CPU hours to finish all the spectral fitting. The model fits to this particular GRB {demonstrate} that the synchrotron model can successfully {explain typical Band function spectra} and a systematic study of a much larger sample is highly encouraged for the future studies.

\acknowledgments
We thank Xiaohong Zhao, Michael J. Burgess, Robert Preece, Linda S. Sparke and Colleen Wilson-Hodge for helpful comments and suggestions. B.B.Z. acknowledges support from GBM through NNM11AA01A/MSFC. B.Z. acknowledges a NASA ADAP program (NNX14AF85G) for support. We acknowledge the use of public data from the Fermi Science Support Center (FSSC).

\end{document}